\documentclass[12pt]{article}
\usepackage{epsfig}
\begin{document}
\def\be{\begin{equation}}\def\ee{\end{equation}}\def\l{\label}
\def\0{\setcounter{equation}{0}}\def\b{\beta}\def\S{\Sigma}\def\C{\cite}
\def\r{\ref}\def\ba{\begin{eqnarray}}\def\ea{\end{eqnarray}}
\def\n{\nonumber}\def\R{\rho}\def\X{\Xi}\def\x{\xi}\def\la{\lambda}
\def\d{\delta}\def\s{\sigma}\def\f{\frac}\def\D{\Delta}\def\pa{\partial}
\def\Th{\Theta}\def\o{\omega}\def\O{\Omega}\def\th{\theta}\def\ga{\gamma}
\def\Ga{\Gamma}\def\t{\times}\def\h{\hat}\def\rar{\rightarrow}
\def\vp{\varphi}\def\inf{\infty}\def\le{\left}\def\ri{\right}
\def\foot{\footnote}\def\ve{\varepsilon}\def\N{\bar{n}(s)}\def\cS{{\cal S}}
\def\k{\kappa}\def\sq{\sqrt{s}}\def\bx{{\bf x}}\def\La{\Lambda}
\def\bb{{\bf b}}\def\bq{{\bf q}}\def\cp{{\cal P}}\def\tg{\tilde{g}}
\def\cf{{\cal F}}\def\bN{{\bf N}}\def\Re{{\rm Re}}\def\Im{{\rm Im}}
\def\K{{\cal K}}\def\ep{\epsilon}\def\cd{{\cal d}}\def\co{\hat{\cal
O}} \def\j{{\h j}}\def\e{{\h e}}\def\F{{\bar{F}}}\def\cn{{\cal N}}
\def\P{\Phi}\def\p{\phi}\def\cd{\cdot}\def\L{{\cal L}}\def\U{{\cal U}}
\def\Z{{\cal Z}}\def\ep{\epsilon}\def\a{\alpha}\def\ru{{\rm u}}
\def\vep{\varepsilon}

\title
{\Large\bf Yang-Mills Fields Quantization in the Factor Space}
\author{ J.Manjavidze$^{a)}$, A.Sissakian$^{b)}$}
\maketitle

\vskip 0.8cm
\begin{abstract}
The perturbation theory over inverse interaction constant $1/g$ is
constructed for Yang-Mills theory. It is shown that the new
perturbation theory is free from the gauge ghosts and Gribov's
ambiguities, each order over $1/g$ presents the gauge-invariant
quantity. It is remarkable that offered perturbation theory did
not contain divergences, at least in the vector fields sector, and
no renormalization procedure is necessary for it.
\end{abstract}

PACS numbers: 02.30.Cj, 03.65.Db, 02.40.Vh, 31.15.Kb
\newpage
\section{Introduction}

The perturbation theory for (3+1)-dimensional Yang-Mills field
theory in vicinity of the extremum $\ru^a_\mu(x)$ of the action
will be described $^{1}$. It is our first publication in this
field and it seems reasonable to define from the very beginning
the level of its completeness.  Namely, we would like to show
that, in opposite to the ordinary perturbative QCD (pQCD), the
offered theory may be used at arbitrary distances.  Accordingly,
the theory is free from divergences at least in the vector fields
sector.  Besides the perturbation theory is operate with
transparently gauge invariant quantities and no ghosts and Gribov
ambiguities would hinder the computations.

We will realize the perturbation theory in the factor space ${\cal
G/H}$, where ${\cal G}$ is the symmetry group of theory and ${\cal
H}$ is the symmetry of $\ru^a_\mu(x)$.  Introductory notes for
this formalism was given in $^{2}$.

The usefulness of such choice follows from homogeneity and
isotropy of ${\cal G/H}$ in the semiclassical approximation. The
developed perturbation theory is formulated for description of the
violating these property quantum excitations.  One may note that
we offer the realization of perturbation theory in terms of the
action-angle type variables. As an example one may have in mind
the factor space $^{3}$
\be
W_G={O(4,2)\t G}/{O(4)\t O(2)}, \l{o}\ee where $G$ is the
non-Abelian gauge group.

The formalism will be demonstrated for simplest quantity, - the
vacuum-into-vacuum transition amplitude $$
\Z(\ru)=<vac;\ru|vac;\ru>, $$ along the path $\ru_{\mu}^a(x)$.
Moreover, following to the idea that the calculation should be
adjusted to the experiments ability $^{2}$, we will restrict
ourselves calculating only the modulo squire $$
\cn(\ru)=|<vac;\ru|vac;\ru>|^2=|\Z(\ru)|^2, $$ since, being the
unmeasurable quantity, the phase of $\Z(\ru)$ is not important
from physical point of view $^{4}$ (it is the principle of
`minimal necessity' in our terminology).

This quantity  $\cn(\ru)$ would normalize the observables and is
equal to squire of the volume of ${\cal G/H}$, see $^{5}$. So, it
defines a number of expected on the trajectory $\ru_{\mu a}(x)$
degrees of freedom, i.e.  $(\ln\cn(\ru))/2$ is proportional to the
dimension of ${\cal G/H}$.  In the example (\r{o}):
\be
\dim W_G=\dim G+8 \l{o'}\ee since the $O(4)\t O(2)$-invariant
solution $\ru_{\mu a}(x)$ breaks both the gauge and the spatial
symmetries. Last one includes the translational and spatial
conformal transformations $^{3}$.

Having in consideration the probability-like quantity $\cn(\ru)$,
one can include into formalism the total probabilities
conservation principle (see $^{2}$, where the role of unitarity
condition in formation of quantum dynamics is described in
details). So, one may prove that if we postulate the path-integral
representation for $\Z(\ru)$, see (\r{0i}) for scalar field case,
and take into account the $S$-matrix unitarity condition then, if
the canonical perturbation series exist (at least in Borel sense),
$\cn$ has the following strict path-integral representation:
\be
\cn=e^{-i\K(je)}\int DM_j(A) e^{-2iU(A,e)}. \l{o1}\ee In this
expression $\K(je)$ acts as the differential operator of the
auxiliary variables $j_{a\mu}$ and $e_{a\mu}$ at $j_{a\mu}=
e_{a\mu}=0$, see (\r{m2}) and (\r{f7}), and the expansion of
$\exp\{-i\K\}$ generates the perturbation series.  The functional
$U(A,e)$ defines interaction.  It may be expressed through the
input classical action, see (\r{m3}) and (\r{f8}).  The main point
of our consideration is the differential measure $DM_j$ since it
is $\d$-like:
\be
DM_j(A)=\prod_{a,\mu}\prod_{x} dA^{a\mu}(x)\d\le(\f{\d S(A)} {\d
A_\mu^a(x)}+j^{a\mu}(x)\ri), \l{o3}\ee where $S(A)$ is the
$classical$ Yang-Mills action. Notice that using the Fourier
transform of functional $\d$-function in (\r{o3}), one may easy
find from (\r{o1}) that $\cn(\ru)=|\Z(\ru)|^2$.

The structure  of representation (\r{o1})  did not depends on the
dimension of system, concrete form of the Lagrangian and other
`local' properties of the theory. We will not repeat for this
reason derivation of (\r{o1}) since it is the same as in $^{2}$
(and $^{6}$, where the (1+1)-dimensional exactly integrable field
theory was considered).

Following to the definitions of $\d$-function and operator
$\K(je)$, one should start from the equation:
\be
\f{\d S(A)} {\d A_\mu^a(x)}=0. \l{o4a}\ee So, having a theory on
the $\d$-like measure, we must consider
 $^{2}$ only the {\it strict} solution of Lagrange equation. Notice
that the equation (\r{o4a}) has also the `trivial' solution
$A_\mu^a(x)=0$, with the corresponding factor space $W_0$, $\dim
W_0=\dim G$, where $G$ is the gauge group.  The pQCD presents
expansion around just this `trivial' solution.

Then, if the general position concerning initial data is analyzed,
we should neglect this `trivial' solution since we will assume
that our solution $\ru_{\mu}^a(x)$ live in the factor space of
$\dim ({\cal G/H})>\dim W_0$.  This is a formal reason why the
expansion in vicinity of $\ru_{\mu a}(x)\neq const$ would be
considered. Corresponding realization of the Yang-Mills theory
would be the topological QCD (tQCD).

This selection rule $^{2}$ is our definition of the ground state.
It should be stressed its importance. It says that first of all
one should consider such solution of the Lagrange equation in the
Minkowski space which is live in the factor space ${\cal G}/{\cal
H}$ of highest dimension since, generally speaking, other orbits
are realized on zero measure $^{7}$.

The extraordinary role of the factor space has specific
explanation. At a first glance $\d$-likeness of measure (\r{o3})
solves the problem of path integral calculation. But actually, to
calculate the remaining integral in (\r{o1}), measure (\r{o3})
forces us to search new forms of perturbation theory.  The formal
reason is hidden in inhomogeneity of our Lagrange equation, see
(\r{o3}),
\be
-\f{\d S(A)}{\d A_\mu^a(x)}=j^{a\mu}(x). \l{o4}\ee So, the exact
solutions of this equation are unknown even in the expansion over
$j^{a\mu}(x)$ form if the corresponding homogeneous equation
(\r{o4a}) has nontrivial solution $\ru_{a\mu}(x)\neq$const.

Nevertheless one may try to solve this equation in the form of
some perturbation series, expanding solution over $j^{a\mu}(x)$.
This will lead to the theory which may have a near resemblance of
the canonical one, see e.g. $^{9}$ where the `straight pass'
approximation was considered.

But the canonical perturbation theory for non-Abelian gauge
theories have additional problems.  First of all, the method of
Faddeev-Popov $^{10}$, introduced for separation of dynamical
degrees of freedom from pure gauge ones, in the most cases lead to
the cumbersome perturbation theory with non-unitary ghost fields
Lagrangian $^{11}$.  In the quantum gravity this, at first glance
a technical complication, rise up to fundamental one, see e.g.
$^{12}$.

Then, it was noted that it is impossible to fix the Coulomb gauge
unambiguously for the Yang-Mills potentials of nontrivial topology
$^{13}$.  Moreover, it was shown later that this conclusion did
not depends on the chosen gauge, is general for non-Abelian gauge
theories $^{14}$ if the expansion is builded around the nontrivial
topology gauge orbits $^{15}$.

We will realize another approach to the problem. Namely, we will
consider the mapping into the corresponding to $\ru_{a\mu}=
\ru_{a\mu}(x;\x,\eta,\la_a)$ factor space. Formally the mapping
can be performed since the $\d$-like measure (\r{o3}) defines the
necessary and sufficient set of contributions into the functional
integral.  We will find the explicit form of $\K$, $U$ and $DM_j$
in the ${\cal G/H}$ space. This is our first quantitative result.

Following to the idea formulated in $^{2}$, we will formulate the
transformation in such a way that
$\ru_{a\mu}=\ru_{a\mu}(x;\x,\eta, \la_a)$ would be the generator
of transformation:
\be
\ru_{a\mu}:~A_{a\mu}(x)\to~(\x(t),\eta(t),\la_a(x)), \l{o5}\ee
where the set $(\x,\eta,\la_a)\in {\cal G/H}$ will coincide at
$j^{a\mu}(x)=0$ with integration parameters of eq.(\r{o4a}),
$\la_a(\bx)$ is the gauge phase and the variables $\x$ and $\eta$
are the consequence of the spatial symmetry breaking. For the
example (\r{o}), $\dim(\x+ \eta)=8$. So, the combination of
generators of violated by $\ru_{a\mu}$ subgroup will be taken as
the new quantum variables, instead of the Yang-Mills potentials
$A_{\mu a}$. In other words, just the variables extracted by the
Faddeev-Popov $anzats$ as the `non-physical' ones would be the
dynamical variables of the tQCD.

The problem of definition and farther quantization of the factor
space was solved in $^{2}$. The method consist in formal mapping
into the symplectic phase space $W$ of the arbitrary high
dimension, considering all dynamical variables of extended space
as the $q$-numbers. It is the first step of calculations. Notice
that the transformation always may be done canonically and the
Jacobian of transformation would be equal to one.  For this reason
no ghost fields will appear.

Then the formalism allows to reduce $W$:
\be
W=({\cal G/H})\t R^* \l{o5a}\ee This reduction of $W$ up to ${\cal
G/H}$ is the second step of calculations.  The realized
transformation is singular since $\dim ({\cal G/H})<\dim W$.
Nevertheless we will be able to extract corresponding artifact
infinity equal to the volume of $R^*$ and cancel it by the
normalization.

The prove that the extracted by this way set of $q$-numbers is
necessary and sufficient for quantization of the factor space
${\cal G/H}$ will be crucial for our formalism.  We will find
that:
\be
{\cal G/H}=T^*V\t R, \l{o6}\ee where the quantum degrees of
freedom only are belong to the cotangent symplectic manifold
$T^*V$ $^{16}$ and $R$ is the $c$-number parameters subspace. The
direct product (\r{o6}) means that we will be able to isolate the
quantum degrees of freedom from classical ones.  So, it will be
shown that $\la_a\in R$.

We will find that each order of the tQCD perturbation theory is
transparently gauge invariant. This result seems natural since the
gauge invariant quantity, the `probability' $\cn(\ru)$, is
calculated. Therefore, there will not be a necessity to fix the
gauge and, therefore, no `copies' of Gribov $^{13}$ would arise.
Moreover, it will be shown that no unphysical singularities
connected to the Gribov's ambiguity $^{17}$ would occur in the
formalism.  This is our second quantitative result.

It is not hard to show, see also $^{2}$, that developed
perturbation theory in the ${\cal G/H}$ space presents expansion
over $1/g$, where $g$ is the interaction constant, and does not
contain the terms $\sim g^n$, with $n>0$. Such type of
perturbation theory, over $1/g$, presents a definite problem from
ordinary renormalization procedures point of view.

Indeed, the ordinary quantum field theory scheme assumes the
multiplicative renormalization of the interaction constant: the
renormalized constant $g_R=Z^{1/2}g< \infty$ and the
renormalization factor $Z=\infty$ because of the ultraviolet
divergences. Then, having the expansion over $1/g$, we come to
evident contradiction: it is impossible to have the {\it infinite
multiplicative} renormalizations in expansions over $g$ and over
$1/g$ simultaneously. For this reason this question would be
considered in more details in our approach.  We will show that our
perturbation theory would not contain the divergences and the
problem with renormalization would not arise.  This is our third
result.

It should be noted here that this results have been proposed to be
obtained in $^{18}$ to distinguish the quantization on the factor
space, but now this is done for complete perturbation theory.
However to mention is that the quantitative progress was achieved
taking into account the unitarity condition.

It was mentioned in $^{2}$ that our perturbation theory, over
$1/g$, is dual to ordinary one, over $g$ $^{19}$. So, we may
realize the expansion or over $g$, or over $1/g$, and the choice
is defined only by the convenience. If the states counted by the
expansion over $g$ and over $1/g$ belong to orthogonal Hilbert
spaces $^{20}$ then should not be any connections among terms of
both expansion
 $^{2}$, only the result of summation of series should coincide.  For
this reason our formalism did not hides the contradiction:  the
expansion over $g$ may contain divergences and it needs the
renormalization, but the expansion over $1/g$ may be divergences
free and no renormalizations would be necessary in it $^{21}$.

In the chosen way of calculations even the notion of $interacting$
gluons in the Yang-Mills theory would disappeared (as well as the
pQCD Feynman diagrams). Yet, we can not exclude the real
(mass-shell) particles (gluons) emission $^{22}$ on the to-day
level of understanding of abilities of our formalism and,
therefore, we can not prove that the states counted in the
expansion over $g$ and over $1/g$ belong to the orthogonal Hilbert
space.  So, we will leave unsolved the problem of colored quanta
emission since the question of confinement demands the more
careful analysis.

The paper is organized as follows. Considering the solutions of
Yang-Mills equation, one may use the $ansatz$ $^{23}$:
\be
A^a_{\mu}(x)=\eta^a_{\mu\nu}\pa^\nu\ln \vp(x), \l{o7}\ee where
$\eta^a_{\mu\nu}$ are the real matrices. This ansatz reduce the
Yang-Mills equation to the form $^{3}$:
\be
\pa^2\vp+\k\vp^3=0, \l{o8}\ee where $\k$ is the integration
constant. So,  in Sec.2 we will formulate the ideology of mapping
into the simpler factor space $W=O(4,2)/O(4)\t O(2)$ for scalar
$O(4,2)$-invariant field theory with the action:
\be
S(\vp)=\int d^4x\le(\f{1}{2}(\pa_\mu\vp)^2-\f{\k}{4}\vp^4\ri).
\l{o9}\ee

In Sec.3 we will formulate the tQCD in the ${\cal G/H}$ factor.

\section{Scalar conformally invariant field theory}

\subsection{Definitions}

We concentrate an attention in present section on calculation of
$|\Z(\ru)|^2$, where
\be
\Z=\int D\vp e^{iS(\vp)} \l{0i}\ee and $S(\vp)$ is the action
defined in (\r{o9})

As was explained, the integral
\be
\cn\equiv|\Z|^2=e^{-i\K(je)}\int DM_j(\vp,\pi)e^{-2iU(\vp,e)}
\l{m1}\ee will be analyzed instead of (\r{0i}). Here
\be
2\K(je)=\Re\int_{C_+}dx \f{\d}{\d j(x)}\f{\d}{\d e(x)}\equiv
\Re\int_{C_+}dx\j(x)\e(x). \l{m2}\ee At the very end of
calculations one should take the auxiliary variables $j$ and $e$
equal to zero. The interactions are introduced by the functional
$$ -2U(\vp,e)=S_{C_+}(\vp+e)-S_{C_-}(\vp-e)- 2\Re\int_{C_+} d^4x
e\f{\d S(\vp)}{\d\vp}= $$\be
=2\k\Re\int_{C_+}dx\vp(x)e^3(x)+O(\ep). \l{m3}\ee The complex time
formalism of Mills $^{24}$ was used and $S_{C_\pm}$ is the action
defined on the complex time contour $C_\pm$. For sake of
definiteness, we will use the complex time contours
\be
C_\pm:~t\to t\pm i\ep,~\ep\to+0,~|t|\leq\infty. \l{m6}\ee

Let $\vp_\pm$ are the fields on the $C_\pm$ branches of the Mills
time contour and let $\pa C_\pm$ is the boundary of this branches.
It was assumed the `periodic' (closed-path $^{6}$) boundary
condition:
\be
\vp_+(t\in \pa C_+)=\vp_-(t\in \pa C_-). \l{m6a}\ee when the
representation (\r{m1}) was derived. This boundary condition
should be maintained in the factor space.

Notice that considering the theory with Lagrangian (\r{o9}), one
may write $U(\vp,e)$ in the following equivalent form (with
$O(\ep)$ accuracy) :
\be
3!U(\vp,e)=-\int d^4x e(x)^3\f{\d^3}{\d\vp(x)^3}S(\vp) = -\int
d^4x \le\{e(x)\f{\d}{\d\vp(x)}\ri\}^3S(\vp), \l{m3'}\ee This
representation is useful for investigation of the perturbation
theory symmetry properties. The indication that the contribution
belongs to the Mills time contour was not shown in (\r{m3'}) since
it was assumed that, for instance,
\be
\f{\d j(t\in C_a)}{\d j(t'\in C_b)}=\d_{ab}\d(t-t'),~a,b=+,-.
\l{m3a}\ee For this reason it is sufficient to indicate the branch
of Mills contour only in the definition of the operator (\r{m2}).

We will consider the `phase space' motion:
\be
DM_j(\vp,\pi)=\prod_x d\vp(x)d\pi(x) \d\le(\dot{\vp}(x)-\f{\d
H_j}{\d\pi(x)}\ri) \d\le(\dot{\pi}(x)+\f{\d H_j}{\d\vp(x)}\ri).
\l{m4}\ee It is important that the formalism involves the $total$
Hamiltonian
\be
H_j=\int d^3x\le[\f{1}{2}\pi^2+\f{1}{2}(\nabla\vp)^2+
\f{\k}{4}\vp^4-j\vp\ri] \l{m5}\ee and the last term $\sim j\vp$
may be interpreted as the time dependent energy of random quantum
excitations. It is evident that we may find the measure (\r{o6})
if the first $\d$-function in (\r{m4}) is used to calculate the
integral over $\pi$. Thus, the representation (\r{m1}), with the
measure (\r{m4}), may be considered as the `first-order'
formalism.

This ends the definition of the field theory on the Dirac measure.

\subsection{Mapping into the factor space}

Having a theory defined on the $\d$-like measure, arbitrary
transformations are easily available. We will start from general
situation introducing $N$ fields $\{\x(x),\eta(x)\}_N$, $N$ is
arbitrary.

To perform the transformation:
\be
(\vp(x),\pi(x))\to\{\x(x),\eta(x)\}_N \l{i1}\ee one should insert
\be
1=\f{1}{\D(\vp,\pi)}\int D\x D\eta
\prod_x\d(F_\x(\vp,\pi;\x,\eta))
\prod_x\d(F_\eta(\vp,\pi;\x,\eta)) \l{i1'}\ee into the integral
(\r{m1}). The functional $\d$-function $\prod_x\d$ has following
properties: $$ \int DX \prod_x\d(X(x))=1,~ $$\be \int DX
\prod_x\d(\pa_\mu X(x))= \int \prod_x dX(x)\d(\pa_\mu X(x))=
\int\prod_{x\neq x_\mu}dX_{(\mu)}(x). \l{i0}\ee Here
$X_{(\mu)}(x)$ is the solution of equation $\pa_\mu X(x)=0$, i.e.
is the arbitrary, including constant, $x_\mu$ independent
function.

Having the measure (\r{m4}) and inserting the unit (\r{i1'}) into
(\r{m1}) the integrals of type: $$ \int D\x D\eta D\vp D\pi
\D^{-1}(\vp,\pi)\t $$\be \t\prod
\d(F_\x(\vp,\pi;\x,\eta))\d(F_\eta(\vp,\pi;\x,\eta))
\d\le(\dot{\vp}-\f{\d H_j}{\d\pi}\ri) \d\le(\dot{\pi}+\f{\d
H_j}{\d\vp}\ri) \l{i2g}\ee would appear. Notice that the $(\dim\x
+\dim\eta)=N$ was chosen arbitrary.

It is important that both measures in (\r{i2g}), over $(\x,\eta)$
and over $(\vp,\pi)$, are $\d$-like.  This allows to change order
of integration and integrate firstly over $\vp$ and $\pi$. It is
natural, at first glance, to use for this purpose last two
$\d$-functions.  Then first ones will define the constraint. This
scheme may restore the WKB perturbation theory, if the unite
(\r{i1'}) is reduced to the Faddeev-Popov $ansatz$ $^{2}$.  But if
the first two $\d$-functions of (\r{i2g}) are used to calculate
the integrals over $\vp$ and $\pi$, we perform transformation to
the new dynamical variables $(\x,\eta)$. Then the last two
$\d$-function will give the dynamical equations for $(\x,\eta)$.
Both ways of computation would give the same result since one may
use arbitrary $\d$-functions.

Thus, we wish to use the fact that the $\d$-like measure defines a
complete set of contributions. Moreover, as follows from (\r{m1})
and (\r{m2}), the quantum perturbations, both in the $(\vp,\pi)\in
V$ and $(\x,\eta)\in W$ spaces, would be generated by the same
operator $\exp\{-i\K(je)\}$ and the interactions in both above
cases are described by the same functional $U(u,e)$.  This
circumstances allows to describe the $quantum$ dynamics in terms
of new variables.

Then, if the `phase space flow' $(u,p)$ belongs to the manifold
${\cal G/H}$ completely, we should be able to `restore' it through
the $(u,p)$ flow. This is our key idea. We will see that this
order of computation, inverse to ordinary one $^{25}$, mostly
natural for us since it allows to start transformation from mostly
general variables $(\x,\eta)\in W$.

Following space-time local realization of the algebraic equations
was offered in $^{2,6}$: $$
F_\x(\vp,\pi;\x,\eta)=\vp(x)-u(x;\x(x),\eta(x))=0, $$\be
F_\eta(\vp,\pi;\x,\eta)=\pi(x)-p(x;\x(x),\eta(x))=0, \l{j1}\ee
where $u=u(x;\x(x),\eta(x))$, $p=p(x;\x(x),\eta(x))$  are some
$compound$ functions. We will assume that this functions would be
defined in accordance with our choice of ${\cal G/H}$. The
equalities (\r{j1}) can be satisfied for arbitrary given
$u(x;\x(x),\eta(x))$, $p(x;\x(x),\eta(x))$ and arbitrary $N$ since
integration over all $\vp(x)$ and $\pi(x)$ is assumed.

Therefore, integral in (\r{i1'}) is not equal to zero since,
generally speaking, always exist. The result of integration in
(\r{i1'}) is denoted by $\D(\vp,\pi)$ and in this sense the
equality (\r{i1'}) is satisfied identically. The additional
constraints for $u(x;\x,\eta)$ and $p(x;\x,\eta)$ will be offered
later.

We will specify (\r{j1}) adding the condition that the time
dependence is hidden in $\x(y,t)$ and $\eta(y,t)$, $x=(y,t)$,
$\dim (y)=3$.  Thus, we would use, instead of (\r{j1}), the
equations:
\be
\vp(y,t)=u(y;\x(y,t),\eta(y,t)),~ \pi(y,t)=p(y;\x(y,t),\eta(y,t)).
\l{j3}\ee In other aspects the functions $u(y;\x,\eta)$,
$p(y;\x,\eta)$ for the time being are arbitrary.  Notice that
offered additional condition is evident since $(u,p)$ would belong
to ${\cal G/H}$ completely.  But, nevertheless, we will examine it
$^{27}$. Notice also the noncovariantness of equalities (\r{j3}).
This is a consequence of necessity to use the Hamiltonian
formalism $^{2}$.

The integration measures in (\r{i2g}) over $\x(y,t)$ and
$\eta(y,t)$ are defined on the total Mills time contour
$C=C_++C_-$:
\be
\int_{C}dt=\int_{C_++C_-}dt=\int_{C_+}dt+\int_{C_-}dt, \l{j3'}\ee
and the integration should be performed with boundary condition
(\r{m6a}):
\be
u(;\x(,t\in\pa C_+),\eta(,t\in\pa C_+))= u(;\x(,t\in\pa
C_-),\eta(,t\in\pa C_-)). \l{j3a}\ee Depending to the topology of
the trajectory $u(;\x,\eta)$, this boundary condition may lead to
nontrivial consequences.

The mapping (\r{j3}) is generated by the function $u$:
\be
u:~(\vp,\pi)\to (\x,\eta) \l{j3''}\ee since the `first-order'
formalism is considered. It is important also to note that this
transformation did not conserves the dimension:
\be
\dim(\vp,\pi)(y,t)\neq\dim(\x,\eta)(y,t) \l{1j2}\ee since
$(\x,\eta)\in {\cal G/H}$ and $(\vp,\pi)\in V$.

{\it{\bf Proposition I}. Jacobian of transformation of the
$\d$-like measure always can be done equal to one.}

Using first two $\d$-functions in (\r{i2g}) to perform integration
over $(\vp,\pi)$ the Jacobian of the transformation (\r{j3''})
takes the form:
\be
J=\f{1}{\D(u,p)}\prod_{y,t} \d\le(\dot{u}(y;\x,\eta)-\f{\d
H_j(u,p)}{\d p(y;\x,\eta)}\ri) \d\le(\dot{p}(y;\x,\eta)+\f{\d
H_j(u,p)}{\d u(y;\x,\eta)}\ri), \l{j4}\ee where the definitions
(\r{j1}) and (\r{j3}) was used. Notice that $\D=\D(u,p)$, as a
result of integration over $\vp$ and $\pi$.

We should dioganalize arguments of remaining $\d$-functions. For
this purpose following trick will be used $^{2}$. So, for
instance, $$ \d\le(\dot{u}-\f{\d H_j}{\d p}\ri)=
\d\le(u_\x\cdot\dot{\x}+u_\eta\cdot\dot{\eta}-\f{\d H_j}{\d
p}\ri)= $$$$ =\d\le(u_\x\cdot\le\{\dot{\x}-\f{\d
h_j}{\d\eta}\ri\}+ u_\eta\cdot\le\{\dot{\eta}+\f{\d
h_j}{\d\x}\ri\} +u_\x\cdot\f{\d h_j}{\d\eta}-u_\eta\cdot\f{\d
h_j}{\d\x} -\f{\d H_j}{\d p}\ri), $$ where  $u_X\equiv\pa u/\pa
X$, $X=\x,\eta$ and $h_j=h_j(\x,\eta)$ is the auxiliary
functional. Let us choose it by the equality:
\be
u_\x\cdot\f{\d h_j}{\d\eta}-u_\eta\cdot\f{\d h_j}{\d\x}- \f{\d
H_j}{\d p}= \f{\pa u}{\pa\x}\cdot\f{\d h_j}{\d\eta}- \f{\pa
u}{\pa\eta}\cdot\f{\d h_j}{\d\x}- \f{\d H_j}{\d p}=\{u,h_j\}-\f{\d
H_j}{\d p}=0, \l{j5}\ee where $\{,\}$ is the Poisson bracket. The
scalar product means that the sets $\{\x\}$ and $\{\eta\}$ was
ordered in such a way that the Poisson bracket would be well
defined. This ordering always possible iff $W$ is the symplectic
manifold.

Then, if (\r{j5}) is satisfied, $$ \d\le(\dot{u}-\f{\d H_j}{\d
p}\ri)= \d\le(u_\x\le\{\dot{\x}-\f{\d h_j}{\d\eta}\ri\}+
u_\eta\le\{\dot{\eta}+\f{\d h_j}{\d\x}\ri\}\ri), $$ The analogous
expression one may find for second $\d$-function: $$
\d\le(\dot{p}+\f{\d H_j}{\d u}\ri)= \d\le(p_\x\le\{\dot{\x}-\f{\d
h_j}{\d\eta}\ri\}+ p_\eta\le\{\dot{\eta}+\f{\d
h_j}{\d\x}\ri\}\ri), $$ and $h_j$ and $p$ should obey additional
to (\r{j5}) equality:
\be
\{p,h_j\}+\f{\d H_j}{\d u}=0. \l{j6}\ee On this stage two equality
(\r{j5}) and (\r{j6}) are the equations for functions
$u(;\x,\eta)$, $p(;\x,\eta)$ and $h_j(\x,\eta)$. Thus, being
vague, this mechanism of mapping is able to endure more
constraints.

Using the ordinary property of $\d$-function: $$ \d(a-b)=\int dc
\d(c-a)\d(c-b), $$ we can write that: $$ J(\x,\eta)=\f{1}{\D(u,p)}
\int D\x'D\eta'\prod_x
\d(u_\x\cdot\x'+u_\eta\cdot\eta')\d(p_\x\cdot\x'+p_\eta\cdot\eta')
\t$$\be\t \d\le(\x'-\le\{\dot{\x}-\f{\d h_j}{\d\eta}\ri\}\ri)
\d\le(\eta'-\le\{\dot{\eta}+\f{\d h_j}{\d\x}\ri\}\ri). \l{j7'}\ee

Let us assume that the functional integral $\D(u,p)$ may be
written in the form: $$ \D(u,p)= $$$$ =\int D\x' D\eta'
\prod_{y,t} \d(\vp(y,t)-u(y;\x+\x',\eta+\eta'))
\d(\pi(y,t)-p(y;\x+\x',\eta+\eta'))= $$\be =\int D\x' D\eta'
\prod_{y,t} \d(u_\x\x'+u_\eta\eta')\d(p_\x\x'+p_\eta\eta')\neq0
\l{j2}\ee This is possible since the functions $\vp(y,t)$ and
$\pi(y,t)$ was chosen in such a way that the equalities (\r{j1})
are satisfied. The inequality (\r{j2}) excludes the degeneracy.
For this reason only $\x'=\eta'=0$ are essential in the integral
(\r{j2}).

In result the determinant $\D(u,p)$ is canceled identically:
\be
DM_j(\x,\eta)=\prod_{y,t}d\x(y,t)\eta(y,t)
\d\le(\dot{\x}(y,t)-\f{\d h_j}{\d\eta(y,t)}\ri)
\d\le(\dot{\eta}(y,t)+\f{\d h_j}{\d\x(y,t)}\ri) \l{j7}\ee since
one may leave arbitrary pare of $\d$-functions in (\r{j7'}) and
$\x'=\eta'=0$ are essential.  Therefore, because of cancelation of
the functional determinants our perturbation theory would be free
from the ghost fields. This considerably simplifies the described
formalism. Notice that the equalities (\r{j5}), (\r{j6}) and
(\r{j2}) should be satisfied to have this result.

The transformed measure (\r{j7}) depends on the auxiliary
functional $h_j=h_j(\x,\eta)$, defined by the equalities (\r{j5})
and (\r{j6}). So, choosing $arbitrary$ $u(;\x,\eta)$ and
$p(;\x,\eta)$ with the property (\r{j2}), one may find $h_j$ from
(\r{j5}) and (\r{j6}), and then (\r{j7}) would be the transformed
measure.

Therefore, mapping (\r{j3''}) based on the equations (\r{j5}) and
(\r{j6}) admits one more equation for $u(;\x,\eta)$, $p(;\x,\eta)$
and $h_j(\x,\eta)$. We will consider following example in present
paper. One may note from (\r{j7}) that $h_j$ has a meaning of
transformed Hamiltonian of new equations:
\be
\dot{\x}(y,t)=\f{\d h_j(\x,\eta)}{\d\eta(y,t)},~
\dot{\eta}(y,t)=-\f{\d h_j(\x,\eta)}{\d\x(y,t)}. \l{j9}\ee

{\it {\bf Proposition II}. If
\be
h_j(\x,\eta)=H_j(u,p), \l{j10}\ee then the Poisson equations
(\r{j5}), (\r{j6}) would define the `phase space flow' $(u,p)$}.

Indeed, having in mind (\r{j3}),
\be
\dot{u}=u_\x\dot{\x}+u_\eta\dot{\eta}= u_\x\f{\d
h_j}{\d\eta}-u_\eta\f{\d h_j}{\d\x}= \{u,h_j\}=\f{\d H_j}{\d p},
\l{j11}\ee where (\r{j9}) and then (\r{j5}) were used. The same
equation one may find for $p$:
\be
\dot p=p_\x\dot{\x}+p_\eta\dot{\eta}= p_\x\f{\d
h_j}{\d\eta}-p_\eta\f{\d h_j}{\d\x}=\{p,h_j\}= -\f{\d H_j}{\d u}.
\l{j12}\ee Therefore, having (\r{j10}) the equations (\r{j5}),
(\r{j6}), simultaneously with (\r{j9}), are equal to the
Hamiltonian equations (\r{j11}) and (\r{j12}).  Notice also that
in this case the time dependence actually should be hidden into
$\x$ and $\eta$.

It should be stressed also that as follows (\r{j11}) and (\r{j12})
fixed by (\r{j5}), (\r{j6}) and completed by (\r{j10}) and
(\r{j2}) transformations unique in those respects that other
`type' of mapping would lead to `unnatural', much more complicate,
formalism.

Having (\r{j5}), (\r{j6}), (\r{j10}) and taking into account
(\r{j2}), we get to the `overdetermined' system of constraints,
which may be inconsistent. The Coulomb problem gives quantum
mechanical example of such system $^{2}$. At all evidence, the
$O(4)\t O(2)$-invariant solution did not obey (\r{j2}) also. On
other hand, if we reject (\r{j2}) then the determinant $\D(u,p)$
is not canceled and the formalism would contain the ghosts.

\subsection{Structure of dual perturbation theory}

The problem of mapping for the degenerate case was solved in
$^{2}$. It was assumed that one may `softly' take off the
degeneracy, i.e. exist some parameter $\vep\to0$ which regulates
the strength of degeneracy breaking and at $\vep=0$ we have the
degenerate limit $^{26}$. Following proposition will be important
in this connection.

{\it {\bf Proposition III}. The quantum perturbation conserves the
topology of phase space flow.}

Indeed, notice that the equations (\r{j5}) and  (\r{j6}) should be
satisfied for arbitrary $j(y,t)$. Let us consider the consequence
of this proposition. Remembering (\r{m5}), and using the
definition (\r{j10}), we find that (\r{j5}) at $j=0$ gives
equality: $$ \le\{u_\x p_\eta-u_\eta p_\x-1\ri\}\f{\d H}{\d
p(y,t)} =\le\{u_\eta u_\x-u_\x u_\eta\ri\}\f{\pa H}{\pa u(y,t)},
~~H=\le. H_j\ri|_{j=0}. $$ Here $u$ and $p$ are the compound
functions of $\x=\x(y,t)$ and $\eta=\eta(y,t)$. This equality is
identically satisfied if the space-time local Poison brackets:
\be
\{u(y,t),p(y,t)\}=1,~~\{u(y,t),u(y,t)\}=0 \l{j14}\ee are
satisfied. The equation (\r{j6}) at $j=0$ adds following
conditions:
\be
\{u(y,t),p(y,t)\}=1,~~\{p(y,t),p(y,t)\}=0 \l{j15}\ee It is not
hard to see that the higher orders over $j$ did not give new
conditions, i.e. the Poison algebra, completed by (\r{j10}), is
closed.  In other words, the quantum perturbations conserve the
topology $^{28}$ of the phase space flow.

The proposition {\bf III} means that the quantum perturbations
would not alter the structure of $u=u(;\x,\eta)$ and
$p=p(;\x,\eta)$ and they are solution of $classical$ (homogeneous)
equations:
\be
\{u(y;\x,\eta),h(\x,\eta)\}=\f{\d H(u,p)}{\d p(y;\x,\eta)},~
\{p(y;\x,\eta),h(\x,\eta)\}=-\f{\d H(u,p)}{\d u(y;\x,\eta)}.
\l{j16}\ee The $j$ dependence is defined by the equations
({\r{j9}) and is confined completely in $\x$ and $\eta$ only.

So, we may start from a theory with generalized Hamiltonian:
\be
h_j(\x,\eta)=H_j(u,p)+\vep\tilde{H}_j(u,p), \l{j10a}\ee where the
additive term $\sim\vep\to 0$. This proposition means that the
`direct' mechanism of degeneracy breaking is considered $^{26}$
and the Hamiltonian $h_j(\x,\eta)$ may be chosen in such a way
that some of $derivatives$ over auxiliary (artificial) fields
$\x'$ and $\eta'$ have a property:
\be
u_{\x'}\sim u_{\eta'}\sim p_{\x'}\sim
p_{\eta'}\sim\vep\to0,~~(\x',\eta')\in R^*. \l{j17}\ee This is
enough to formulate conserving the phase space volume
transformation of quantum theory.

Thus, we start from the variables $(\x,\eta)\in W$ and scalar
functions $u=u(y;\x,\eta)$, $p=p(y;\x,\eta)$. They should obey the
inequality (\r{j2}) and define the functional $h_j(\x,\eta)$
through the equations (\r{j5}), (\r{j6}).  This allows to cancel
the determinant $\D(u,p)$. Then we extract the auxiliary variables
$\x'$ and $\eta'$ assuming (\r{j17}).  This will allow to exclude
the auxiliary variables and should reduce the system to physical
one. The physical content of this procedure was described in
$^{2}$.

Following property of the perturbation theory in the $W$ space
will be used to realize this program of reduction. In result of
our mapping the integral $\cn$ takes the form:
\be
\cn(\ru)=e^{-i\K(je)}\int DM_j(\x,\eta)e^{-2iU(u,e)}, \l{mm1}\ee
where $DM_j(\x,\eta)$ is defined in (\r{j7}). Notice that in this
expression $U$ depends on $u=u(y;\x,\eta)$.

It was shown in $^{2}$ that the mapped representation (\r{mm1})
allows to split the `quantum force' $j(y,t)$ and corresponding
`virtual field' $e(y,t)$ on the projection on the axes of $W$.  It
is easy to find the result of this procedure:
\be
2\K(je)=\Re\int_{C_+} d^3xdt\le\{\j_\x(y,t)\cdot \e_\x(y,t)+
\j_\eta(y,t)\cdot e_\eta(y,t)\ri\} \l{mm2}\ee and
\be
e=e_\x\cdot\f{\pa u}{\pa\eta}-e_\eta\cdot\f{\pa u}{\pa\x}.
\l{mm3}\ee The hat symbol in (\r{mm2}) means the derivative over
corresponding quantity. At the very end of calculation one should
take $j_X=e_X=0$, $X=(\x,\eta)$. The scalar product means
summation over all components of $\x$ and $\eta$.

Inserting (\r{mm3}) into (\r{m3'}) one can find that $$
-3!U(u,e)=\int d^3xdt \le\{e_\x\cdot\f{\pa u}{\pa\eta}\f{\d}{\d
u}- e_\eta\cdot\f{\pa u}{\pa\x}\f{\d}{\d u}\ri\}^3S(u)= $$\be=
\int d^3xdt\le\{e_\x\cdot\f{\pa u}{\pa\eta}\f{\pa}{\pa u}-
e_\eta\cdot\f{\pa u}{\pa\x}\f{\pa}{\pa u}\ri\}^3\L(u), \l{mm3'}\ee
where $\L(u)$ is the Lagrangian density. This shows that the
interaction functional $U(u,e)$ has the symmetry properties of the
Lagrangian density.

Formally new perturbation generating operator (\r{mm2}) gives the
same perturbation series, but with the rearranged sequence of
terms, i.e. the splitting of $j$ did not change the `convergence'
of the perturbation series (over $1/\k$ since $u\sim1/\sqrt{k}$).
At the same time, this splitting of the source $j$ is useful since
allows to analyze the excitation of each degree of freedom, i.e.
of components of the phase space flow along the axis of $W$,
independently.

Noting that $e_X$, $X=\x,\eta$, is conjugate to $j_X$, it is easy
to conclude that the action of the operator (\r{mm2}) leads to the
operator $$%
\le\{\f{\d}{\d j_\x}\cdot\f{\pa u}{\pa\eta}\f{\pa}{\pa u}-
\f{\d}{\d j_\eta}\cdot\f{\pa u}{\pa\x}\f{\pa}{\pa
u}\ri\}\sim\{\j\wedge\hat{X}\}. $$ This operator is the invariant
of canonical transformations. If by some reason
$d\o^2_X=\j_X\wedge\hat{X}=0$, then the motion along the $X$-th
axis will be classical. This is the mechanism of reduction of the
quantum degrees of freedom. Firstly this important properties of
our formalism was described in $^2$. We will continue this
question in Sec.2.4.

{\it {\bf Proposition IV}. New fields $\x$ and $\eta$ can not
depend on the coordinate $y$ if the scalar theory is considered},
i.e.
\be
\x=\x(t),~~\eta=\eta(t), \l{j17a}\ee for scalar theory (\r{o9}).

This conclusion follows from proposition {\bf III}. The reason is
that the dynamical problem was divided on two parts.  First part
of the problem consist in solution of the $classical$ equations
(\r{j16}). It defines a structure of the compound functions
$u(y;\x,\eta)$ and $p(y;\x,\eta)$. The second part consist in
definition of the $time$ dependence of $(\x,\eta)$ through the
equations (\r{j9}) and (\r{j10}). Finally, if $(\x,\eta)$ in zero
order over $j(y,t)$ are the $y$ independent parameters, the
quantum perturbations are unable to change this property.

It is noticeable that if $\x=\x(t)$ and $\eta=\eta(t)$ then we
will find from (\r{j5}) and (\r{j6}), instead of (\r{j14}) and
(\r{j15}), the $canonical$ equal-time commutator relations:
\be
\{u(y;\x(t),\eta(t)),p(y';\x(t),\eta(t))\}=\d(y-y'). \l{j18}\ee
Thus, our quantization scheme would restore the canonical one in
the factor space $W$. In this sense the independence of $\x$ and
$\eta$ from $y$ is natural.

Nevertheless it seems useful to demonstrate the proposition {\bf
IV} explicitly.  The elements (\r{mm2}) and (\r{j7}) are used in
Appendix A to demonstrate the reduction:
\be
(\x,\eta)(y,t)\to(\x,\eta)(t). \l{mm5a}\ee This involves reduction
of the operators:
\be
(\j_X,\e_X)(y,t)\to(\j_X,\e_X)(t),~X=\x,\eta. \l{mm5b}\ee The
structure of corresponding perturbation theory is described in
subsequent subsection.

\subsection{Reduction}

Therefore, for considered scalar theory,
\be
2\K(je)=\Re\int_{C_+} dt\le\{\j_\x(t)\cdot \e_\x(t)+
\j_\eta(t)\cdot e_\eta(t)\ri\} \l{mm2'}\ee and
\be
e(y;\x(t),\eta(t))= e_\x(t)\cdot\f{\pa
u(y;\x(t),\eta(t))}{\pa\eta(t)}- e_\eta(t)\cdot\f{\pa
u(y;\x(t),\eta(t))}{\pa\x(t)}. \l{mm3z}\ee

The result of disappearance of the $y$ dependencies in $\x$ and
$\eta$ is reduction of the field-theoretical problem to the
quantum mechanical one.  So, $L(u)=V(\x,\eta)$ play here the role
of the mechanical potential for a particle with the {\it phase
space} coordinate $(\x,\eta)$.

The measure takes the form:
\be
DM_j(\x,\eta)=\prod_{t} d\x(t)d\eta(t)
\d(\dot{\x}(t)-\o_\eta(\x,\eta)-j_\x(t))
\d(\dot{\eta}(t)+\o_\x(\x,\eta)-j_\eta(t)), \l{mm4a}\ee where the
`velocity'
\be
\o_X(\x,\eta)=\f{\pa h(\x,\eta)}{\pa X}. \l{mm5e}\ee

Let us remember now the definition (\r{j17}):
\be
u=u(y;\x(t),\eta(t);\vep\x'(t),\vep\eta'(t)),~~\vep\to0, \l{q2}\ee
where
\be
\dim\x=n,~\dim\eta=m,~\dim(\x+\x')=\dim(\eta+\eta')=N. \l{q2a}\ee
Inserting (\r{q2}) into Lagrangian, we find that:
\be
L(u)=\int d^3x\L(u(y;\x(t),\eta(t)))+O(\vep). \l{q3}\ee We are
able now to define the dimension of $T^*V$ taking
\be
N=\dim({\cal G/H}). \l{q3a}\ee So, $N=8$ for the example (\r{o}).

{\it {\bf Proposition V}. If we have (\r{q2}) and (\r{q2a}) then
\be
\dim T^*V=\min\{n,m\} \l{q2d}\ee }

Let us consider following three possibilities to demonstrate this
proposition.

({\bf a}). $n=m$, $N=2n$.

In this case the interaction functional $U(u,e)$ takes the form:
$$ -3!U(u,e)=\int dt\le\{ \le(e_x\cdot\f{\pa}{\pa\eta}-
e_\eta\cdot\f{\pa}{\pa\x}\ri)_{n}\ri.+ $$$$
+\le.\le(e_{x'}\cdot\f{\pa}{\pa\eta'}-
e_{\eta'}\cdot\f{\pa}{\pa\x'}\ri)_{N-n} \ri\}^3 L(u) =$$\be= \int
dt\le\{\le(e_x\cdot\f{\pa}{\pa\eta}-
e_\eta\cdot\f{\pa}{\pa\x}\ri)_n\ri\}^3L(u), \l{q4}\ee where
(\r{q3}) was used. The index $n$ means that the scalar products
include $n$ terms, and $N$ may be chosen equal to $n$. The measure
$$ DM_j(\x,\eta)= \prod_{t} d^n\x(t)d^n\eta(t)
\d^{(n)}(\dot{\x}-\o_\eta-j_\x) \d^{(n)}(\dot{\eta}+\o_\x-j_\eta).
$$

({\bf b}). $n>m$, $N=n+m$.

In this case
\be
-3!U(u,e)= \int dt\le\{\le(e_x\cdot\f{\pa}{\pa\eta}-
e_\eta\cdot\f{\pa}{\pa\x}\ri)_m+
\le(e_\eta'\cdot\f{\pa}{\pa\x}\ri)_{(n-m)}\ri\}^3V(\x,\eta),
\l{q4b}\ee since $\eta'$ is absent in $V(\x,\eta)$. Therefore,
$e_\eta'$ has only the $(n-m)$ components.

The measure takes the form: $$ DM_j(\x,\eta)= \prod_{t}
d^n\x(t)d^m\eta(t)d^{(n-m)}\eta'(t) $$$$
\d^{(m)}(\dot{\x}-\o_\eta-j_\x) \d^{(m)}(\dot{\eta}+\o_\x-j_\eta)
\d^{(n-m)}(\dot{\x}-j_\x) \d^{(n-m)}(\dot{\eta}'+\o_\x-j_{\eta'})
$$ since $N=(n+m)$. Notice that $\eta'$ is contained only in the
argument of last $\d$-function. For this reason we always can
perform the shift: $\dot{\eta}'\to\dot{\eta}'-\o_\x+j_{\eta'}$. In
result: $$ DM_j(\x,\eta)= $$$$ =\prod_{t} d^n\x(t)\eta^m(t)
\d^{(m)}(\dot{\x}-\o_\eta-j_\x) \d^{(m)}(\dot{\eta}+\o_\x-j_\eta)
\d^{(n-m)}(\dot{\x}-j_\x) \d^{(n-m)}(\dot{\eta}') $$ and the
$j_{\eta'}$ dependence is disappeared. For this reason the
$\j_{\eta'}$ dependence in the operator $\K$ may be omitted.  In
result, $$ 2\K(je)=\Re\int_{C_+} dt\le\{(\j_\x\cdot \e_\x)_m+
(\j_\eta\cdot\e_\eta)_m+ (\j_\x\cdot \e_\x)_{(n-m)}\ri\}. $$ There
is not operator $\e_\eta'$ and, for this reason, one should take
$e_{\eta'}$ equal to zero. Therefore,
\be
-3!U(u,e)= \int dt\le\{e_x\cdot\f{\pa}{\pa\eta}-
e_\eta\cdot\f{\pa}{\pa\x}\ri\}_m^3V(\x,\eta) \l{q4c}\ee and the
$(n-m)$ components of $e_\x$ and $j_\x$ may be taken equal to zero
everywhere:
\be
2\K(je)=\Re\int_{C_+} dt\le\{\j_\x\cdot \e_\x+
\j_\eta\cdot\e_\eta\ri\}_m. \l{q4d}\ee Accordingly,
\be
DM_j(\x,\eta)=dR \prod_{t} d^m\x(t)d^m\eta(t)
\d^{(m)}(\dot{\x}-\o_\eta-j_\x) \d^{(m)}(\dot{\eta}+\o_\x-j_\eta),
\l{q4e}\ee where
\be
dR=d^{(N-2m)}\x(0) \l{q4f}\ee is the element of $R$. The trivial
auxiliary elements was omitted.

The same analyses may be done for the case $n<m$.

In result, assuming that $\eta$ is the `action' variable, $$
\o_\eta=\o(\eta)\equiv\pa h(\eta)/\pa\eta,~\o_\x=0, $$ we can
write:
\be
DM_j(\x,\eta)=dR\prod_{i=1}^{\min\{m,n\}} \prod_{t}
d\x_i(t)\eta_i(t) \d(\dot{\x}_i-\o_i(\eta)-j_{i\x})
\d(\dot{\eta}_i-j_{i\eta}). \l{q6}\ee Therefore,
\be
W=T^*V\t R \l{q8}\ee and $dR$ is the differential measure of the
subspace $R$.

This ends the prove of proposition {\bf V}.

So, the equations for $\x$ and $\eta$ take the form:
\be
\dot{\x}(t)=\o(\eta)+j_\x(t),~ \dot{\eta}(t)=j_\eta(t)
\l{mm13b}\ee The second equation is simply integrable:
\be
\eta(t)=\eta_0+\int dt' g(t- t')j_\eta(t')\equiv\eta_0+\eta_j(t).
\l{mm13c}\ee Inserting this solution into the first equation in
(\r{mm13b}) one may find:
\be
\x(t)=\x_0+\int dt' g(t- t')\o(\eta_0+\eta_j(t'))+\int dt' g(t-
t')j_\x(t')\equiv\x_0+\bar{\o}_j(t)t+\x_j(t), \l{mm13d}\ee where
the abbreviation:
\be
\bar{\o}(t)t=\int dt' g(t- t')\o(\eta_0+\eta_j(t')) \l{mm13e}\ee
was used. The Green function $g(t-t')$ was defined in $^{2}$:
\be
g(t-t')=\Th(t-t'), \l{mm13h}\ee where $\Th(t-t')$ is the step
function with boundary property:
\be
\Th(0)=1. \l{mm13i}\ee

In result,
\be
u=u(y;\x_0+\bar{\o}_j(t)t+\x_j,\eta_0+\eta_j) \l{mm13f}\ee and the
term $$ \sim\f{1}{n!}\{-2i\U(u,j)\}^n=O(\f{1}{\k^n}) $$ gives the
$n$-th order of our perturbation theory over $1/\k$ since
$u=O(1/\sqrt\k)$.

\section{Non-Abelian gauge field theory}

\subsection{Yang-Mills theory on Dirac measure}

The action of considered theory
\be
S(A)=\f{1}{2g}\int d^4x F_{\mu\nu a}(A)F^{\mu\nu}_a(A) \l{f3}\ee
is the $O(4,2)$ invariant and the Yang-Mills fields
\be
F_{\mu\nu a}(A)=\pa_\mu A_{\nu a}-\pa_\nu A_{\mu a}-
C_a^{bc}A_{\mu b}A_{\nu c} \l{f4}\ee are the covariant of
non-Abelian gauge transformations. The gauge group will not be
specified.

We will consider the integral
\be
\cn=e^{-i\K(je)}\int DM_je^{-2iU(A,e)}, \l{f3a}\ee where the
measure
\be
DM_j(A)=\prod_{\mu,a}\prod_{x}dA_\mu^a(x,t)\d(D^{\nu b}_a
F_{\nu\mu b}-j_{\mu a}) \l{f6}\ee is manifestly conformal and
gauge invariant if $j_{\mu a}=0$. The covariant derivative $$
D^{\mu b}_a=\pa^\mu\d^b_a+C^{bc}_a A^\mu_{ c}. $$ The
perturbations generating operator
\be
2\K(je)=\Re\int_{C_+} d^4x\f{\d}{\d j^{\mu}_a(x,t)} \f{\d}{\d
e_{\mu a}(x,t)}\equiv \Re\int_{C_+} d^4x\h{j}_{\mu
a}(x,t)\h{e}^\mu_a(x,t), \l{f7}\ee The auxiliary variables $j_{\mu
a}$ and $e^\mu_a$ should be taken equal to zero at the very end of
calculations. The functional
\be
-2U(A,e)=(S_{C_+}(A+e)-S_{C_-}(A-e)) -2\Re\int_{C_+}d^4x
e^\mu_a(x)\f{\d S (A)}{\d A^\mu_a} +O(\vep) \l{f8}\ee describes
interactions. All above quantities are defined on the Mills time
contours
\be
C_\pm:~t\rar t\pm i\ep,~\ep\rar +0,~|t|\leq\infty. \l{f9}\ee This
gives the rule as to avoid the light-cone singularities solving
the equation:
\be
D^{\nu b}_aF_{\nu\mu b}=j_{\mu a}. \l{f5}\ee One can omit in
(\r{f8}) terms $\sim\ep\to+0$. Therefore, $U(A,e)=O(e^3)$ and may
contain only the odd powers of $e_{a\mu}$. This means that we may
write $U(A,e)$ in the form:
\be
U(A,e)=-\int d^4x\le\{e^\mu_a(x)\f{\d}{\d A^\mu_a(x)}\ri\}^3S(A),
\l{f8a}\ee see (\r{m3'}).

\subsection{First-order formalism}

The noncovariant first order formulation in terms of the
`electric' field
\be
E^i_a=F^{i0}_a, \l{p1}\ee presents introduction into the necessary
for us Hamiltonian description.  The action in this term has the
form
\be
S_{C_\pm}(A,F)=\f{1}{g}\int_{C_\pm} d^4x
\le\{\dot{\bf{A}}_a\cd{\bf{E}}_a+ \f{1}{2}({\bf E}_a^2+{\bf
B}_a^2({\bf A})) -A_{0a}({\bf D}\cd{\bf E})_a\ri\}, \l{p2}\ee
where the `magnetic' field
\be
B_{ia}({\bf A})=({rot}{\bf A})_{ia}+\f{1}{2}\ve_{ijk}[A_j,A_k]_a
\l{p2a}\ee is not the independent quantity and was introduce to
shorten the formulae.  Notice that $A_{0a}$ did not contain the
conjugate pare and the action $S$ is linear over it.

The measure (\r{f6}) may be written in the first-order formalism
representation ($d{\bf{A}}_a=\prod_idA_{ia}$): $$
DM_j({\bf{A}},{\bf{P}})=\prod_{a,i}\prod_{x} d{\bf{A}}_{ai}(x)
d{\bf{P}}_{ai}(x) \d({\bf D}_a^b\cd{\bf P}_b)\t $$\be\t
\d\le(\dot{{\bf{P}}}_a(x)+\f{\d H_j({\bf{A}},{\bf{P}})}
{\d{\bf{A}}_a(x)}\ri) \d\le(\dot{{\bf{A}}}_a(x)-\f{\d
H_j({\bf{A}},{\bf{P}})} {\d{\bf{P}}_a(x)}\ri), \l{f10a}\ee where
$H_j({\bf{A}},{\bf{P}})$ is the total Hamiltonian:
\be
H_j=\f{1}{2g}\int d^3x\le({\bf P}^2_a+{\bf B}_a^2({\bf A})\ri)
+\int d^3x {\bf j}_a {\bf A}_a, \l{a4}\ee
${\bf{P}}_a(x)\equiv{\bf{E}}_a(x)$ is the conjugate to
${\bf{A}}_a(x)$ momentum and ${\bf B}_a(\bf{A})$ was defined in
(\r{p2a}). We may introduce into $DM_j$ additional $\d$-function:
\be
\prod_{a}\prod_{x} \d({\bf B}^i_a-({\rm rot}{\bf A})^i_a-
\f{1}{2}\ve^i_{jk}[A^j,A^k]_a). \l{p2b}\ee Then the Hamiltonian in
(\r{a4}) becomes symmetric over electric ${\bf E}_a$ and magnetic
${\bf B}_a$ fields.

Notice that the first $\d$-function in (\r{f10a}) is the
consequence of linearity of the action over $A_{0a}$. The time
component $A_{0a}$ has the meaning of Lagrange multiplier for the
Gauss law:
\be
{\bf D}_a^b\cd{\bf P}_b=0. \l{f10'}\ee It should be stressed that
there is not equation for the time component $A_{0a}$. Moreover,
the $A_{0a}$ dependence was completely disappeared from formalism
since the interaction functional $U(A,e)$ is defined by the third
derivative over $A_{\mu a}$, see (\r{f8a}).

\subsection{Mapping into the factor space}

The measure (\r{f10a}) is not physical since it contains three
(for given $a$) vector potential ${\bf{A}}_a(x)$. To exclude the
unphysical degree of freedom, the gauge fixing Faddeev-Popov
$ansatz$ is oftenly used. But we will consider, as was described
in previous section, another approach.

We will introduce the functional $$ \D(A,P)= $$\be =\int D\x D\eta
\prod_{a}\d\le({\bf{A}}_a(x)-{\bf{u}}_a(x;\x(x),\eta(x)\ri)
\d\le({\bf{P}}_a(x)-{\bf{p}}_a(x;\x(x),\eta(x)\ri) \l{f11}\ee to
realize the transformation
\be
u:~ (A,P)_a(x)\to(\x,\eta)(x), \l{i3}\ee to the compound vector
functions $({\bf u,p})_a(x;\x(x),\eta(x))$ of the space-time local
parameters $(\x,\eta)(x)$. It is assumed that $\D\neq0$.

Performing transformation (\r{i3}), we find: $$
DM_j(\x,\eta)=\f{1}{\D_c(u)}\prod_{a}\prod_{x}{d\x d\eta d\la_a
dq_a} \d({\bf D}_a^b\cd{\bf p}_b) \t$$\be\t
\d\le(\dot{{\bf{u}}}_a(x)-\f{\d H_j}{\d{\bf{p}}_a(x)}\ri)
\d\le(\dot{{\bf{p}}}_a(x)+\f{\d H_j}{\d{\bf{u}}_a(x)}\ri).
\l{f16}\ee Here the gauge phase $\la_a$ and conjugate to it $q_a$
was extracted from the set of variables $\x$ and $\eta$.

Using the result of previous section, one may dioganalize
arguments of $\d$-functions. In result: $$
DM_j(\x,\eta,\la,Q)=\prod_{x,t,a}d\x d\eta d\la dq \d\le({\bf
D}_a^b({\bf u})\cd{\bf p}_b\ri) \d\le(\dot{\la}_a-\f{\d h_j}{\d
q_a}\ri) \d\le(\dot{q}_a+\f{\d h_j}{\d \la_a}\ri)\t $$\be
\t\d\le(\dot{\x}-\f{\pa h_j}{\pa\eta}\ri) \d\le(\dot{\eta}+\f{\pa
h_j}{\pa\x}\ri). \l{f17}\ee The equality (\r{f17}) is hold iff
$h_j$ is defined by Poisson equations (for the 3-vectors given
${\bf u}_a$ and ${\bf p}_a$):
\be
\{{\bf u}_a(x),h_j\}=\f{\d H_j}{\d{\bf{p}}_a(x)},~ \{{\bf
p}_a(x),h_j\}=-\f{\d H_j}{\d{\bf{u}}_a(x)} \l{f18}\ee considering
$(\x,\eta)$ and $(\la,q)$ in the Poisson brackets as the
canonically conjugate pares.

If we add to (\r{f18}) one more equation:
\be
h_j(\x,\eta,\la,q)=H_j({\bf u}_a,{\bf p}_a) \l{f19}\ee then, as
was shown in previous section, ${\bf u}_a$ and ${\bf p}_a$ should
be solution of incident equations, assuming that (\r{f18}) are
hold on the measure (\r{f17}). Then
\be
{\bf D}_a^b(u)\cd {\bf p}_b\equiv0 \l{g2}\ee since ${\bf p}_b$ is
the solution of eq.(\r{f18}) at arbitrary $j_{\mu a}$. This
remarkable result is the consequence of mapping into the invariant
space ${\cal G/H}$ to which the classical flow belongs completely.
Therefore, corresponding $\d$-function in (\r{p11}) gives
identically $$ \prod_{x}\d(0). $$ This infinite factor should be
canceled by normalization and will not be mentioned later. Note
that the formalism contains one sources ${\bf j}_a$ conjugate to
the coordinates ${\bf u}_a$ only, see (\r{f19}) and (\r{a4}).

So, described mapping gives the measure: $$
DM_j(\x,\eta,\la,Q)=\prod_{x,t;a}d\la_adq_ad\x d\eta
\d\le(\dot{\la}_a\ri)\d\le(\dot{q}_a+\f{\d h_j}{\d \la_a}\ri)
$$\be \d\le(\dot{\x}-\f{\pa h_j}{\pa\eta}\ri)
\d\le(\dot{\eta}+\f{\pa h_j}{\pa\x}\ri) \l{p5}\ee We have took
into account here that $(u,p)_a$ are $q_a$ independent. The
Hamiltonian $h_j$ is defined by eq.(\r{f19}):
\be
2gh_j=\int d^3x \le(p^2_a+ {\bf B}^2_a(u)\ri) + \int d^3x {\bf
j}_a {\bf u}_a\equiv h+J. \l{p6}\ee where $h$ is the unperturbated
by ${\bf j}_a$ Hamiltonian.

Helping the proposition {\bf $V$}, we can exclude the $q_a$
dependence:
\be
DM_j(\x,\eta,\la)=dR\prod_{x;a} d\la_a d\x d\eta \d(\dot{\la}_a)
\d\le(\dot{\x}-\o-j_\x\ri)\d\le(\dot{\eta}-j_\eta\ri) \l{p7}\ee
where the `velocity' $\o={\pa h}/{\pa\eta}.$ The perturbations
generating operator takes the form:
\be
2\K(je)=\int dt\{\j_\x\e_\x+ \j_\eta\e_\eta\}. \l{p9}\ee At the
same time one should replace in (\r{f8}) ${\bf e}_a$ on
\be
{\bf e}_a(x)=e_\x(t)\f{\pa {\bf u}_a(x;\x,\eta,\la)}{\pa\eta(t)}-
e_\eta(t)\f{\pa {\bf u}_a(x;\x,\eta,\la)}{\pa\x(t)}. \l{p10}\ee

As follows from (\r{p7}) we should consider the time independent
gauge transformations:
\be
\dot{\la}_a(x)=0. \l{g1}\ee To remove this constraint we should
generalize equation (\r{f18}). So, if we consider the equation:
\be
\{{\bf u}_a(x;\x,\eta,\la),h_j\}=\f{\d H_j}{\d{\bf{p}}_a(x)}-
\O_a(x)\f{\pa{\bf u}(x;\x,\eta,\la)}{\pa\la_a} \l{p14}\ee instead
of first equation in (\r{f18}) then one should replace in (\r{p7})
\be
\prod_{x;a}d\la_a(x)\d(\dot\la_a(x))\rar\prod_{x;a}
d\la_a(x)\d(\dot\la_a(x)-\O_a(x)), \l{p15}\ee where $\O_a(x)$ is
the arbitrary function of $y$ and $t$. This is the mostly general
representation for gauge measure in our formalism.

In result, the main elements of quantum Yang-Mills theory in the
${\cal G/H}$ space looks as follows:

(i) The measure
\be
DM_j(\x,\eta,\la)=dR\prod_{x;a}d\la_ad\x d\eta
\d(\dot\la_a(x)-\O_a(x))\d(\dot{\x}-\o-j_\x)\d(\dot{\eta}-j_\eta)
\l{p11}\ee Using the definition (\r{i0}), one may note that $$
\int \prod_{x;a}d\la_a\d(\dot\la_a(x)-\O_a(x)) $$ means
integration over all functions $\la_a(y,t)$ of the arbitrary given
time dependence. At the same time
\be
\f{\int \prod_{x;a}d\la_a\d(\dot\la_a(x)-\O_a(x))}{ \int
\prod_{x;a}d\la_a}\equiv0. \l{p11a}\ee Therefore our normalization
on the gauge group volume differs from ordinary one. But this will
not affect the result since all contributions will be gauge
invariant.

(ii) The quantum perturbations generating operator
\be
2\h{K}({\bf je})=\int dt\{\h{\bf j}_\x\cd\h{\bf e}_\x+ \h{\bf
j}_\eta\cd\h{\bf e}_\eta\} \l{p12}\ee

(iii) The interactions functional $U({\bf u},\bar{e})$ depends on
\be
{\bf e}_a={\bf e}_\x\cdot\f{\pa {\bf u}_a}{\pa\eta}- {\bf
e}_\eta\cdot\f{\pa {\bf u}_a}{\pa\x}. \l{p13}\ee Note the motion
along $\la$ orbits is exactly classical and the dependence of
nondynamical variables was disappeared.

\subsection{Gauge invariance}

We wish to quantize the theory without gauge fixing $ansatz$ and,
therefore, the theory contains three $independent$ potential $\bf
u_{i a}$, $i=1,2,3$ for each color index $a$. We may avoid this
problem with the unphysical degrees of freedom if the theory would
depend only from the gauge-invariant observable quantities: the
color electric, ${\bf E}_a$, and magnetic, ${\bf B}_a$, fields.

{\it {\bf Proposition VI}. Each order over $1/g$ is explicitly
gauge invariant}

The interactions functional $U$ has following explicit form: $$
-3!U({\bf u},{e})=\f{1}{g}\int dx \prod_{k=1}^3 \le\{
e_{a_k}\f{\pa}{\pa u_{a_k}}\ri\} F^{\mu\nu a}F_{\mu\nu a}, $$
where $e_a$ was defined in (\r{p13}). Using this definition, we
find:
\be
-3!U({\bf u},{e})=\int dx \prod_{k=1}^3 \le\{\le[{\bf
e}_\x\cdot\f{\pa {\bf u}_a}{\pa\eta}- {\bf e}_\eta\cdot\f{\pa {\bf
u}_a}{\pa\x}\ri]\f{\pa}{\pa\bf{u}_{a_k}}\ri\} F^{\mu\nu
a}F_{\mu\nu a}. \l{g3a}\ee The summation over repeated indices is
assumed.

Last expression is manifestly gauge invariant since the operator
is singlet of gauge transformations and $F^{\mu\nu a}F_{\mu\nu a}$
is the gauge invariant quantity.

\subsection{Divergences}

The expression (\r{g3a}) may be written in the form:
\be
-3!U({\bf u},\bar{e})=\int dt \prod_{k=1}^3 \le\{\le[{\bf
e}_\x\cdot\f{\pa {\bf u}_a}{\pa\eta}- {\bf e}_\eta\cdot\f{\pa {\bf
u}_a}{\pa\x}\ri]\f{\pa}{\pa\bf{u}_{a_k}}\ri\}{\cal L}(u),
\l{g4}\ee where $$ {\cal L}(u)=\int d^3x F^{\mu\nu a}F_{\mu\nu a}
$$ is the Yang-Mills Lagrangian.

Result of action of the perturbation generating operator gives the
expression:
\be
\cn(u)=\int DM(\x,\eta):e^{-2iU(u,e)}:, \l{g5}\ee where the
operator
\be
-3!(2i)^3\U(u,e)=\int dt \prod_{k=1}^3 \le\{\le[\f{\d}{\d{\bf
j}_\x}\cdot\f{\pa {\bf u}_a}{\pa\eta}- \f{\d}{\d{\bf
j}_\eta}\cdot\f{\pa {\bf
u}_a}{\pa\x}\ri]\f{\pa}{\pa\bf{u}_{a_k}}\ri\}{\cal L}(u),
\l{g6}\ee where $u_(ia)$ depends on the solution of equations:
\be
\dot{\x}-\o(\eta)=j_\x,~~~\dot{\eta}=j_\eta \l{g7}\ee and the
measure is $j_X$, $X=\x,\eta$ independent: $$
DM=dR\prod_a\prod_{y,t}D\la_a\d(\dot{\la}_a-\O)\d(\dot{\x}-\o(\eta))
\d(\dot\eta). $$ Such `shift' is possible since the equations
(\r{g7}) are linear over $j_X$.

We can conclude that if $u_{a\mu}$ is not singular,
\be
|S(u)|<\infty, \l{oo}\ee then the theory did not contain
divergences since the differential operator in (\r{g6}) can not
change convergence of the time integrals. Notice that the $O(4)\t
O(2)$ solution obey this property $^3$.

\section{Conclusion}

It was shown that exist such formulation of the quantum Yang-Mills
theory which is (a) divergences free (at least in the vector
fields sector), (b) did not contain the gauge ghosts and (c) is
sufficiently consistent, i.e.  the quantization scheme is free
from the Gribov ambiguities.

It was shown in $^{2}$ that if $\pa({\cal G/H})$ is the boundary
then the quantum corrections are accumulated on this boundary,
i.e. the intersection $\pa u_{a\mu}\bigcap\pa({\cal G/H})$, where
$\pa u_{a\mu}$ is the flow in the ${\cal G/H}$ coordinate system,
defines the value of quantum corrections. If $\pa
u_{a\mu}\bigcap\pa({\cal G/H})=0$ then the semiclassical
approximation is exact. This is the crucial property of our
topological QCD.

For this reason the tQCD seems attractive and the question, may it
take the place of pQCD is seems important. The experimentally
examined consequences of the tQCD would be extremely interesting
and they will be investigated in the first place.

Being convergent, the exactness of estimation of the measurables
in tQCD should be higher then in the `logarithmic' pQCD.
Moreover, the convergence means that the main contributions are
accumulated on the large distances. This property is typical for
hadron physics. Therefore, the main point of our future
publications would be the prediction of the small-scale effects,
where we can compare our approach with pQCD.

\vskip 0.4cm {\large \bf Acknowledgement}

We would like to thank our colleagues in the Lab. of Theor. Phys.
of JINR and especially V.Kadyshevski for fruitful interest to
described technique and underling idea. We are thankful to
V.Ter-Antonian for
 discussions. One of us (J.M.) would like to thank N.Russakovich
and G.Chelkov for kind hospitality at the Lab. of Nucl. Probl.
(JINR).

\vspace{0.25in}

\renewcommand{\theequation}{a.\arabic{equation}}
\appendix\section{Appendix. Reduction of the space degrees of
freedom}\0

Action of the operator $\exp\{-i\K\}$ leads to the expression:
\be
\cn(u)=\int DM_j(\x,\eta):e^{-2i\U(u,j)}:. \l{mm1a}\ee where
\be
-3!(2i)^3\U(u,j)=\int d^3xdt \le\{\le[\f{\d}{\d{\bf
j}_\x}\cdot\f{\pa {\bf u}_a}{\pa\eta}- \f{\d}{\d{\bf
j}_\eta}\cdot\f{\pa {\bf u}_a}{\pa\x}\ri]\f{\pa}{\pa\bf{u}_{a_k}}
\ri\}\L(u) \l{mm1b}\ee and the colons in ({\r{mm1a}) mean the
`normal product', when the variational derivatives over $j_X$ in
the expansion of $\exp\{-2i\U(u,j)\}$ stay to the left of all
functions.

The measure $$ DM_j(\x,\eta)=\prod_{y,t} d\x d\eta
\d(\dot{\x}-\o_\eta-j_\x)\d(\dot{\eta}+\o_\x-j_\eta) $$ Then, to
calculate the remaining integral in (\r{mm1a}), one should find
solution of inhomogeneous equations:
\be
\dot{\x}(y,t)-\o_\eta(y,t;\x,\eta)=j_\x(y,t),~~~
\dot{\eta}(y,t)+\o_\x(y,t;\x,\eta)=j_\eta(y,t), \l{mm11}\ee where
$$ \o_X(y,t;\x,\eta)=\d h(\x,\eta)/\d X(y,t). $$

As follows from (\r{mm1b}), if some operators $\j_{X'}$ over the
`auxiliary' variable $X'$ did not contained in $\U(u,j)$ then the
auxiliary variables $X'$ should obey the homogeneous, classical,
equations, with $j_{X'}=0$ in the right hand side.

The solutions of inhomogeneous equation (\r{mm11}) will be
searched expanding over $j_X$: $$ \x(y,t)=\x^0(y,t)+\int
d^4x'\x^1_\x(y,t;y',t')j_\x(y',t')+ $$$$ +\int
d^4x'\x^1_\eta(y,t;y',t')j_\eta(y',t')+..., $$$$
\eta(y,t)=\eta^0(y,t)+ \int
d^4x'\eta^1_\eta(y,t;y',t')j_\eta(y',t') + $$\be +\int
d^4x'\eta^1_\x(y,t;y',t')j_\x(y',t')+... \l{mm12}\ee So, the
equations:
\be
\dot{\x}^0(y,t)=\o_\eta(y,t;\x^0,\eta^0),~
\dot{\eta}^0(y,t)=-\o_\x(y,t;\x^0,\eta^0) \l{mm6}\ee should be
solved in the lowest order over $j_X$. The function
$u(y;\x(y,t),\eta(y,t))$ should obey the `boundary' property:
\be
\le.u(y;\x(y,t),\eta(y,t))\ri|_{j=0}=u(y;\x^0,\eta^0)=
\ru(y,t;\x_0,\eta_0) \l{mm7}\ee where $\x_0$ and $\eta_0$ are the
integration constants of the Lagrange equation (\r{o8}). The
equality (\r{mm7}) defines the starting set of the necessary
variables $\x$ and $\eta$. Notice that, as follows from
proposition {\bf III}, the quantum perturbations should not change
this set.

Let us distinguish the variables $\x\in {\cal G/H}$ by the
equality:
\be
\le.\f{\d}{\d\x}h\ri|_{j_X=0}=0. \l{mm8}\ee This assumes that the
set $\eta$ can be expressed through the set conserved generators.
In the example (\r{o}), they are the generators of translation and
special conformal transformation. Notice, that the proposition
{\bf III} mens that the quantum perturbations did not alter this
definition.

Inserting (\r{mm8}) into (\r{mm6}) we find at $j_X=0$ the
equations:
\be
\dot{\x}^0(y,t)=\o_\eta(\eta^0)\equiv\o(\eta^0),~
\dot{\eta}^0(y,t)=0. \l{mm9}\ee The functions with arbitrary $y$
dependence may satisfy this equations. Using solution of this
equations:
\be
\x^0(y,t)=\o(\eta^0)t+\x_0,~~~\eta^0(y,t)=\eta_0, \l{mm10}\ee
where $\x_0$ and $\eta_0$ are the integration constants, we will
see that the dependence on $y$ in (\r{mm7}) did not play any role
because of the degeneracy over $y$. For this reason we will put
out the $y$ dependence in $\x^0$ and $\eta^0$.

It is not hard to show that the degeneracy over $y$ will conserved
in arbitrary order over $j_X$. Indeed, inserting the expansions
(\r{mm12}) into the equations (\r{mm11}), we find in the first
order over $j_\x$: $$ \pa_t\x^1_\x(y,t;y',t')-\x^1_\x(y,t;y',t')
\le.\f{\d^2 h(\x,\eta)}{\d \x(y',t')\d\x(y,t)}\ri|_{j=0}- $$$$
-\x^1_\eta(y,t;y',t') \le.\f{\d^2 h(\x,\eta)}{\d
\eta(y',t')\d\x(y,t)}\ri|_{j=0}= \d(y-y')\d(t-t'). $$ Notice that
$$ \le.\f{\d h(\x,\eta)}{\d\x(y,t)}\ri|_{j=0}= \f{\d }{\d\x(y,t)}
\le\{\le.h(\x,\eta)\ri|_{j=0}\ri\}=0, $$ where (\r{mm8}) was used.
Therefore, the equation for $\x^1_\x$ have a structure:
\be
\dot{\x}^1_\x(y,t;y',t')=\d(y-y')\d(t-t'), \l{mm13}\ee where the
boundary conditions (\r{mm10}) was applied. Notice that this
equation is linear.

Inserting the solution of equation (\r{mm13}):
\be
{\x}^1_\x(y,t;y',t')=\d(y-y')g(t-t'), \l{mm13a}\ee where $g(t-t')$
is the Green function defined in $^{2}$, into (\r{mm12}), we find
the term $$ \sim\int dt'g(t-t')j_\x(y,t'). $$ So, the $y$
dependence is contained in the auxiliary source $j_\x$ only. For
this reason it can not play dynamical role. The same phenomena one
can observe considering other terms in the decomposition
(\r{mm12}).

Therefore, admitting that the  quantum perturbations switched on
adiabatically, i.e. may be taken into account perturbatively, and
for this reason are unable to change the topology of the classical
trajectory $u(y;\x,\eta)$, the proposition {\bf III}, one may
conclude that it is enough to take $\x=\x(t)$ and $\eta=\eta(t)$
in the considered scalar theory.

\end{document}